\documentclass[a4paper]{jpconf}%
\usepackage{graphicx}
\usepackage{amsmath}
\usepackage{amsfonts}
\usepackage{amssymb}

\begin{document}

\title{Neutrino masses, dark matter and baryon asymmetry of the Universe}
\author{Amine Ahriche$^{(1)}$ and Salah Nasri$^{(2)}$}

\address{$^{(1)}$ Department of Physics, University of Jijel, PB 98 Ouled Aissa,
DZ-18000 Jijel, Algeria.\\
$^{(2)}$ Physics Department, UAE University, POB 17551, Al Ain,
United Arab Emirates.}

\ead{aahriche@ictp.it, snasri@uaeu.ac.ae}

\begin{abstract}
In this work, we try to explain the neutrino mass and mixing data radiatively
at three-loop by extending the standard model (SM) with two charged singlet
scalars and three right handed (RH) neutrinos. Here, the lightest RH neutrino
is a dark matter candidate that gives a relic density in agreement with the
recent Planck data, the model can be consistent with the neutrino oscillation
data, lepton flavor violating processes, the electroweak phase transition can
be strongly first order; and the charged scalars may enhance the branching
ratio $h\rightarrow\gamma\gamma$, where as $h\rightarrow\gamma Z$ get can get
few percent suppression. We also discuss the phenomenological implications of
the RH neutrinos at the collider.

\end{abstract}

\section{Introduction}

The Standard Model (SM) of particle physics has been very successful in
describing nature at the weak scale, however, there are many unexplained
puzzles left, that implies going beyond SM. Three concrete evidences for
Physics beyond SM are: (i) non zero neutrino masses, (ii) the existence of
dark matter (DM), and (iii) the observation of matter anti matter asymmetry of
the universe. However, most of the SM extensions make no attempt to address
these three puzzles within the same framework. For instance, a popular
extension of the SM, is introducing very heavy right-handed (RH) neutrinos
($m_{N}\geq10^{8}~\mathrm{GeV}$), where small neutrino masses are generated
via the see-saw mechanism \cite{seesaw}, and the BAU is produced via
leptogenesis \cite{FY}. Unfortunately, a RH neutrino heavier than $10^{7} $
\textrm{GeV}, decouples from the effective low energy theory, and can not be
tested at collider experiments.

A small neutrino mass can be generated radiatively, where the famous example
is the so-called Zee model \cite{Zee}. In Zee-model, the solar mixing angle
comes out to be close to maximal, which is excluded by the solar neutrino
oscillation data \cite{solnu}. This problem is circumvented in models where
neutrinos are induced at two loops \cite{Babu} or three loops \cite{KNT, AKS}.
One of the advantages of this class of models is that all the mass scales are
in the TeV or sub-TeV range, which makes it possible for them to be tested at
future colliders. In Ref. \cite{KNT}, the SM was extended with two charged
$SU(2)_{L}$ singlet scalars and one RH neutrino field, $N$, where a $Z_{2}$
symmetry was imposed to forbid the Dirac neutrino mass terms at tree level
\cite{KNT}. After the electroweak symmetry breaking, tiny neutrino masses are
naturally generated at three loops due to the high loop suppression. A
consequence of the $Z_{2}$ symmetry and the field content of the model, $N$ is
$Z_{2}$-odd, and thus guaranteed to be stable, which makes it a good DM
candidate. In Ref. \cite{Keung}, the authors considered extending the fermion
sector of the SM with two RH neutrinos, in order for it to be consistent with
the neutrino oscillation data, and they studied also its phenomenological implications.

In \cite{kntJCAP}, we calculated the three loop neutrino masses exactly, as
compared to the approximate expression derived in \cite{KNT}. We have shown
that in order to satisfy the recent experimental bound on the lepton flavor
violating (LFV) process such as $\mu\rightarrow e\gamma$ \cite{LFV}; and the
anomalous magnetic moment of the muon \cite{pdg}, one must have three
generations of RH neutrinos. Taking into account the neutrino oscillation data
and the LFV constraints, we show that the lightest RH neutrino can account for
the DM abundance with masses lighter than 225 \textrm{GeV}. The presence of
the charged scalars in this model will affect the Higgs decay process
$h\rightarrow\gamma\gamma$ and can lead to an enhancement with respect to the
SM, where as $h\rightarrow\gamma Z$ is slightly reduced. In this model, we
find that a strongly electroweak phase transition can be achieved with a Higgs
mass of $\simeq125$ \textrm{GeV} as measured at the LHC \cite{ATLAS,CMS}.

This letter is organized as follows. In the next section we present
the model, and discuss the constraints from the LFV processes. In
section III, we study the relic density of the lightest RH neutrino
and different DM features in this model. Section IV is devoted to
the study of the electroweak phase transition combined with the
effect of the presence of extra charged scalars on the Higgs decay
channels $h\rightarrow\gamma\gamma$\ and $h\rightarrow \gamma Z$. In
section V, we discuss the phenomenological implications of the RH
neutrinos at the ILC. Finally we conclude in section VI.

\section{Neutrino Mass \& Mixing}

\begin{figure}[t]
\begin{centering}
\includegraphics[width=5cm,height=2cm]{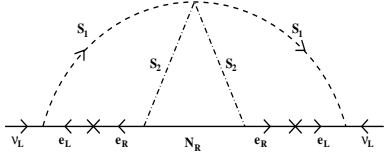}
\par\end{centering}
\caption{\textit{The three-loop diagram that generates the neutrino mass.}}%
\label{diag}%
\end{figure}

Our model is just the SM extended by three RH neutrinos, $N_{i}$, and two
charged singlet scalars, $S_{1}^{\pm}$ and $S_{2}^{\pm}$. In addition, we
impose a discrete $Z_{2}$ symmetry on the model, under which $\{S_{2}%
,N_{i}\}\rightarrow\{-S_{2},-N_{i}\}$, and all other fields are even. The
Lagrangian reads
\begin{equation}
\mathcal{L}=\mathcal{L}_{SM}+\{f_{\alpha\beta}L_{\alpha}^{T}Ci\tau_{2}%
L_{\beta}S_{1}^{+}+g_{i\alpha}N_{i}S_{2}^{+}\ell_{\alpha R}+\tfrac{1}%
{2}m_{N_{i}}N_{i}^{C}N_{i}+h.c\}-V(\Phi,S_{1},S_{2}),\label{L}%
\end{equation}
where $L_{\alpha}$ is the left-handed lepton doublet, $C$ is the charge
conjugation matrix, $f_{\alpha\beta}$ are Yukawa couplings which are
antisymmetric in the generation indices $\alpha$ and $\beta$, $m_{N_{i}}$\ are
the Majorana RH neutrino masses; and $V(\Phi,S_{1},S_{2})$ is the tree-level
scalar potential which is given by
\begin{align}
V(\Phi,S_{i})  &  =\lambda\left(  \left\vert \Phi\right\vert ^{2}\right)
^{2}-\mu^{2}\left\vert \Phi\right\vert ^{2}+%
{\textstyle\sum_{i}}
\{m_{i}^{2}S_{i}^{\ast}S_{i}+\lambda_{i}S_{i}^{\ast}S_{i}\left\vert
\Phi\right\vert ^{2}+\frac{\eta_{i}}{2}\left(  S_{i}^{\ast}S_{i}\right)
^{2}\}\nonumber\\
&  +\left\{  \lambda_{s}S_{1}S_{1}S_{2}^{\ast}S_{2}^{\ast}+h.c\right\}
,\label{nu-mass}%
\end{align}
where\ $\Phi$ is the SM Higgs doublet. The $Z_{2}$ symmetry imposed on the
Lagrangian implies:\newline(1) if $N_{1}$ is the lightest particle among
$N_{2},N_{3},S_{1}$ and $S_{2}$, then it would be stable, and hence it is a
candidate for dark matter. Moreover, $N_{i}$ will be pair produced and
subsequently decay into $N_{1}$ (or to $N_{2}$ and then to $N_{1}$) and a pair
(or two pairs) of charged leptons. We will discuss its phenomenology later.
(2) The Dirac neutrino mass term is forbidden at all levels of the
perturbation theory, and Majorana neutrinos masses are generated radiatively
at three-loops, as shown in Fig. \ref{diag}-left.

The exact estimation of the diagram in Fig. \ref{diag} leads to
\cite{kntJCAP}
\begin{equation}
(M_{\nu})_{\alpha\beta}=\frac{\lambda_{s}m_{\ell_{i}}m_{\ell_{k}}}{\left(
4\pi^{2}\right)  ^{3}m_{S_{2}}}f_{\alpha i}f_{\beta k}g_{ij}g_{kj}F\left(
m_{N_{j}}^{2}/m_{S_{2}}^{2},m_{S_{1}}^{2}/m_{S_{2}}^{2}\right)
,\label{nu-mass-1}%
\end{equation}
where $\rho,\kappa,j(i,k)$ are flavor (eigenstates) indices, and the function
$F$ is a loop integral given in (A.8) in \cite{kntJCAP}, which was
approximated to one in the original work \cite{KNT}. In this model, the
radiatively generated neutrino masses are directly proportional to the charged
leptons and RH neutrino masses as shown in (\ref{nu-mass-1}) unlike the
conventional seesaw mechanism. The neutrino mass matrix elements
(\ref{nu-mass-1}) should fit the experimental values
\begin{equation}
(M_{\nu})_{\alpha\beta}=[U\cdot diag(m_{1},m_{2},m_{3})\cdot U^{T}%
]_{\alpha\beta},\label{comp}%
\end{equation}
where $U$ is the Pontecorvo-Maki-Nakawaga-Sakata (PMNS) mixing matrix
\cite{PMNS}, where the mixing angles are given by the experimental allowed
values for $s_{12}^{2}=0.320_{-0.017}^{+0.016}$\textbf{, }$s_{23}%
^{2}=0.43_{-0.03}^{+0.03}$\textbf{, }$s_{13}^{2}=0.025_{-0.003}^{+0.003}$,
with $s_{ij}\equiv\sin(\theta_{ij})$ and $c_{ij}\equiv\cos(\theta_{ij})$, and
the mass differences\textbf{\ }$\left\vert \Delta m_{31}^{2}\right\vert
=2.55_{-0.09}^{+0.06}\times10^{-3}$ \textrm{eV}$^{2}$ and $\Delta m_{21}%
^{2}=7.62_{-0.19}^{+0.19}\times10^{-5}\mathrm{eV}^{2}$ \cite{GF}.

Besides neutrino masses and mixing, the Lagrangian (\ref{L}) induces flavor
violating processes such as $\ell_{\alpha}\rightarrow\gamma\ell_{\beta} $ if
$m_{\ell_{\alpha}}>m_{\ell_{\beta}}$, generated at one loop via the exchange
of both extra charged scalars $S_{i}^{\pm}$. The branching ratio of such
process can be computed as \cite{kntJCAP}
\begin{equation}
B(\ell_{\alpha}\rightarrow\gamma\ell_{\beta})=\frac{\alpha_{em}\upsilon^{4}%
}{384\pi}\left\{  \frac{\left\vert f_{\kappa\alpha}f_{\kappa\beta}^{\ast
}\right\vert ^{2}}{m_{S_{1}}^{4}}+\frac{36}{m_{S_{2}}^{4}}\left\vert
{\sum\limits_{i}}g_{i\alpha}g_{i\beta}^{\ast}F_{2}\left(  \frac{m_{N_{i}}^{2}%
}{m_{S_{2}}^{2}}\right)  \right\vert ^{2}\right\}  ,
\end{equation}
with $\kappa\neq\alpha,\beta$, $\alpha_{em}$ is the fine structure constant
and $F_{2}(x)=(1-6x+3x^{2}+2x^{3}-6x^{2}\ln x)/6(1-x)^{4}$. Another constraint
which the bound on the muon anomalous magnetic moment $\delta a_{\mu}$, that
receives the contribution \cite{kntJCAP}
\begin{equation}
\delta a_{\mu}=\frac{m_{\mu}^{2}}{16\pi^{2}}\left\{  \frac{\left\vert f_{\mu
e}\right\vert ^{2}+\left\vert f_{\mu\tau}\right\vert ^{2}}{6m_{S_{1}}^{2}%
}+\frac{1}{m_{S_{2}}^{2}}{\sum\limits_{i}}\left\vert g_{i\mu}\right\vert
^{2}F_{2}\left(  \frac{m_{N_{i}}^{2}}{m_{S_{2}}^{2}}\right)  \right\}  .
\end{equation}

\begin{figure}[t]
\begin{centering}
\includegraphics[width=7cm,height=5cm]{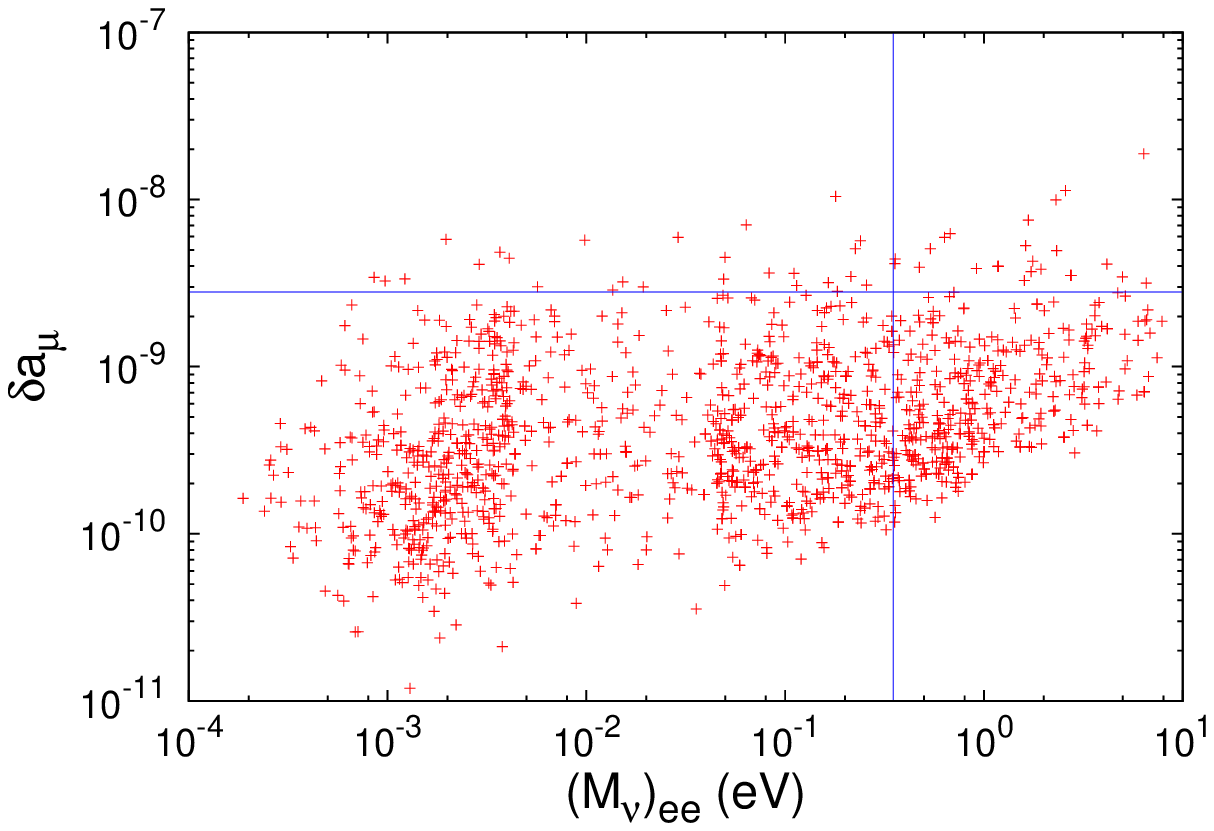} \includegraphics[width=7cm,height=5cm]{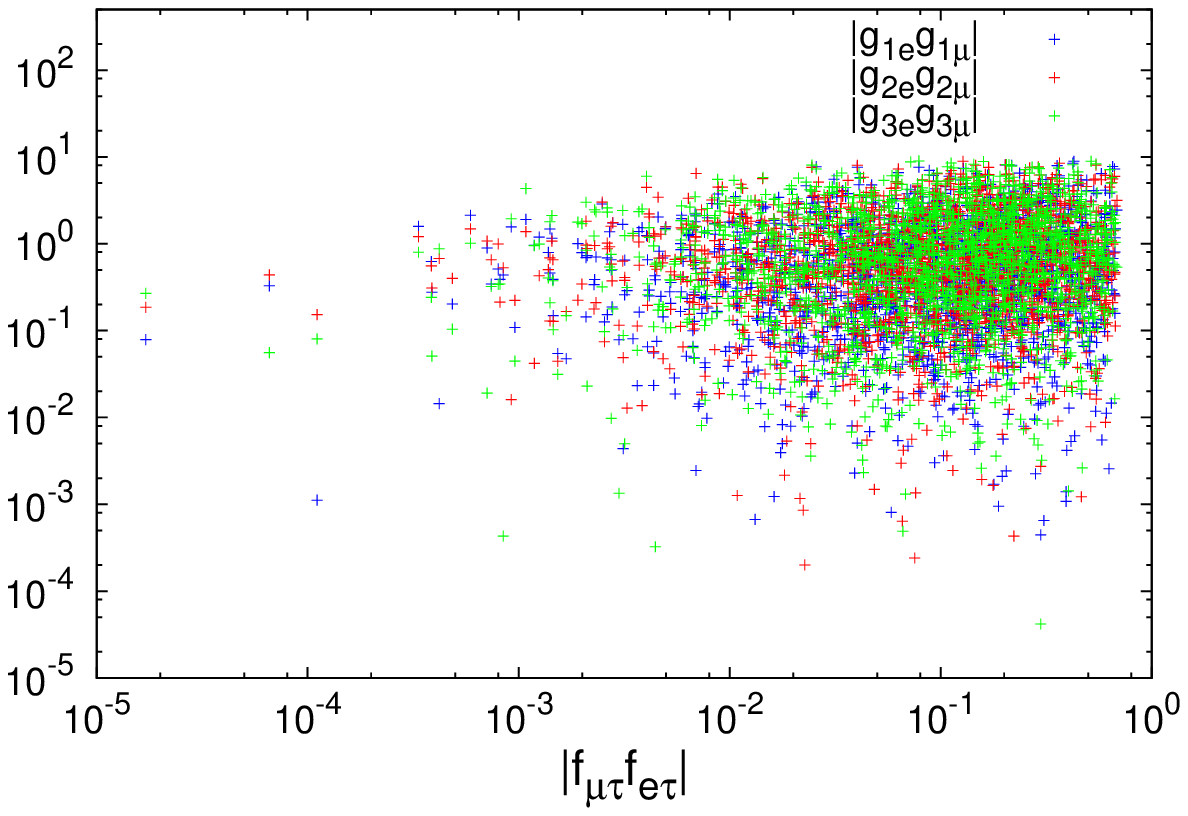}
\par\end{centering}
\caption{\textit{Left: The muon anomalous magnetic moment versus the
$\beta\beta_{0\nu}$ decay effective Majorana $\mathit{(M_{\nu})_{ee}}$. The
blue lines represent their experimental upper bounds. Right: Different
parameters combinations (as absolute values) that are relevant to the LFV
constrain on $B(\mu\rightarrow e\gamma)$, are shown where (\ref{nu-mass-1})
and (\ref{comp}) are matched.}}%
\label{amu}%
\end{figure}

In Fig. \ref{amu}-left, we show a scattered plot of the muon anomalous
magnetic moment versus the $\beta\beta_{0\nu}$ decay effective Majorana mass
$\left(  M_{\nu}\right)  _{ee}$, where the considered upper bound is $\left(
M_{\nu}\right)  _{ee}<0.35~\mathrm{eV}$ \cite{bbo}. In our parameter space
scan, we consider $m_{S_{1,2}}\geq100$ \textrm{GeV}\textbf{; }and demanded
that (\ref{nu-mass-1}) to be consistent with the neutrino oscillation data.
From Fig. \ref{amu}-left, one can see that most of the values of $\left(
M_{\nu}\right)  _{ee}$ that are consistent with the bound on $\delta a_{\mu}$
are lying in the range $10^{-3}$ \textrm{eV} to $\sim$\textrm{eV}.

Fig. \ref{amu}-right gives an idea about the magnitude of the couplings that
satisfy the constraints from LFV processes and the muon anomalous magnetic
moment, and which also are consistent with the neutrino oscillation data. It
is worth noting that when considering just two generations of RH neutrinos
(i.e, $g_{3\alpha}=0$), we find that the bound $B\left(  \mu\rightarrow
e\gamma\right)  <5.7\times10^{-13}$ is violated \cite{LFV}. Therefore, having
three RH neutrinos is necessary for it to be in agreement with the data from
the bounds from LFV processes. Moreover, one has to mention that the bound on
$B\left(  \mu\rightarrow e\gamma\right)  $ makes the parameters space very
constrained. For instance, out of the benchmarks that are in agreement with
the neutrino oscillation data, DM and $\delta a_{\mu}$, only about 15\% of the
points will survive after imposing the $\mu\rightarrow e\gamma$ bound.

\section{Dark Matter}

\begin{figure}[t]
\label{Om}\begin{centering}
\includegraphics[width=7cm,height=5cm]{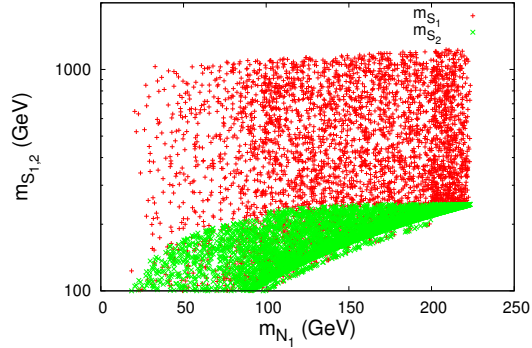}
\par\end{centering}
\caption{\textit{The charged scalar masses $m_{S_{1}}$ (red) and $m_{S_{2}}$
(green) versus the lightest RH neutrino mass, where the consistency with the
neutrino data, LFV constraints and the DM relic density have been imposed.}}%
\label{msmn}%
\end{figure}

As mentioned above, the lightest RH neutrino $N_{1}$ is stable, and could be
the DM candidate. In the case of hierarchical RH neutrino mass spectrum, we
can safely neglect the effect of $N_{2}$ and $N_{3}$ on $N_{1}$ density. The
$N_{1}$ number density get depleted through the annihilation process
$N_{1}N_{1}\rightarrow\ell_{\alpha}\ell_{\beta}$ via the $t$-channel exchange
of $S_{2}^{\pm}$. After estimating the Feynman diagrams, squaring, summing and
averaging over the spin states, we find that in the non-relativistic limit,
the total annihilation cross section is given by
\begin{equation}
\sigma_{N_{1}N_{1}}\upsilon_{r}\simeq\sum_{\alpha,\beta}|g_{1\alpha}g_{1\beta
}^{\ast}|^{2}\frac{m_{N_{1}}^{2}\left(  m_{S_{2}}^{4}+m_{N_{1}}^{4}\right)
}{48\pi\left(  m_{S_{2}}^{2}+m_{N_{1}}^{2}\right)  ^{4}}\upsilon_{r}%
^{2},\label{sig11}%
\end{equation}
with $\upsilon_{r}$ is the relative velocity between the annihilation $N_{1}%
$'s. As the temperature of the universe drops below the freeze-out temperature
$T_{f}\sim m_{N_{1}}/{25}$, the annihilation rate becomes smaller than the
expansion rate (the Hubble parameter) of the universe, and the $N_{1}$'s start
to decouple from the thermal bath. The relic density after the decoupling can
be obtained by solving the Boltzmann equation, and it is approximately given
by \cite{kntJCAP}%
\begin{equation}
\Omega_{N_{1}}h^{2}\simeq\frac{1.28\times10^{-2}}{\sum_{\alpha,\beta
}|g_{1\alpha}g_{1\beta}^{\ast}|^{2}}\left(  \frac{m_{N_{1}}}{135~\mathrm{GeV}%
}\right)  ^{2}\frac{\left(  1+m_{S_{2}}^{2}/m_{N_{1}}^{2}\right)  ^{4}%
}{1+m_{S_{2}}^{4}/m_{N_{1}}^{4}},\label{Omh2}%
\end{equation}
In Fig. \ref{msmn}, we plot the allowed mass range $(m_{N_{1}},m_{S_{i}})$
plane that give the observed dark matter relic density \cite{Planck}. As seen
in the figure, the neutrino experimental data combined with the relic density
seems to prefer $m_{S_{1}}>m_{S_{2}}$ for large space of parameters. However,
the masses of both the DM and the charged scalar $S_{2}^{\pm}$ can not exceed
$m_{N_{1}}<225$ \textrm{GeV} and $m_{S_{2}}<245$ \textrm{GeV}, respectively.

In computing the relic density in (\ref{Omh2}), we have assumed that there is
a hierarchy between the three right-handed neutrino masses. However, when the
possibility for $N_{2}$ and/or $N_{3}$ being close in mass to $N_{1}$, i.e
$\Delta_{i}=(m_{N_{i}}-m_{N_{1}})/m_{N_{1}}<<1$, is considered, the
coannihilation processes like $N_{1}N_{2,3}\rightarrow\ell_{\alpha}\ell
_{\beta}$ might enhance the relic density values by three if the degeneracy is
around $\Delta_{2}\sim0.1$.

\section{Electroweak Phase Transition \& Higgs Decay}

It is well known that the SM has all the qualitative ingredients for
electroweak baryogenesis, but the amount of matter-antimatter asymmetry
generated is too small. One of the reasons is that the electroweak phase
transition (EWPT) is not strongly first order, which is required to suppress
the sphaleron processes in the broken phase. The strength of the EWPT can be
improved if there are new scalar degrees of freedom around the electroweak
scale coupled to the SM Higgs, which is the case in the model that we are
considering in this paper.

The investigation of the transition dynamics and its strength requires the
precise knowledge of the effective potential of the CP-even scalar fields at
finite temperature \cite{Th}. The one-loop Higgs effective potential is given
in the $\overline{DR}$ scheme by
\begin{align}
V_{eff}(h,T) &  =\frac{\lambda}{4!}h^{4}-\frac{\mu^{2}}{2}h^{2}+\sum_{i}%
n_{i}\frac{m_{i}^{4}(h)}{64\pi^{2}}\left(  \ln\left(  \frac{m_{i}^{2}%
(h)}{\Lambda^{2}}\right)  -\frac{3}{2}\right)  ,\nonumber\\
&  +\tfrac{T^{4}}{2\pi^{2}}\sum_{i}n_{i}J_{B,F}\left(  m_{i}^{2}/T^{2}\right)
+V_{ring}(h,T);\label{Veff}\\
V_{ring}(h,T) &  =-\frac{T}{12\pi}\sum\limits_{i}n_{i}\left\{  \tilde{m}%
_{i}^{3}(h,T)-m_{i}^{3}(h)\right\}  ,\label{ring}\\
J_{B,F}\left(  \alpha\right)   &  =\int_{0}^{\infty}x^{2}\log(1\mp\exp
(-\sqrt{x^{2}+\alpha})),\label{JBF}%
\end{align}
where $h=(\sqrt{2}Re(H^{0})-\upsilon)$\ is the real part of the neutral
component in the doublet, $n_{i}$\ are the field multiplicities, $m_{i}%
^{2}(h)$ are the field-dependent mass squared which are given in Appendix B in
\cite{kntJCAP}, and $\Lambda$ is the renormalization scale which we choose to
be the top quark mass. where the summation is performed over the scalar
longitudinal gauge degrees of freedom, and $\tilde{m}_{i}^{2}(h,T)$ are their
thermal masses, which are given in Appendix B. The contribution (\ref{ring})
is obtained by performing the resummation of an infinite class of of infrared
divergent multi-loops, known as the ring (or daisy) diagrams, which describes
a dominant contribution of long distances and gives significant contribution
when massless states appear in a system. It amounts to shifting the
longitudinal gauge boson and the scalar masses obtained by considering only
the first two terms in the effective potential \cite{ring}. This shift in the
thermal masses of longitudinal gauge bosons and not their transverse parts
tends to reduce the strength of the phase transition. The integrals
(\ref{JBF}) is often estimated in the high temperature approximation, however,
in order to take into account the effect of all the (heavy and light) degrees
of freedom, we evaluate them numerically.

In order to generate a baryon asymmetry at the electroweak scale \cite{EWB},
the anomalous violating $B+L$ interactions should be switched-off inside the
nucleated bubbles, which implies the famous condition for a strong first order
phase transition \cite{SFOPT}
\begin{equation}
\upsilon(T_{c})/T_{c}>1,\label{v/t}%
\end{equation}
where $T_{c}$ is the critical temperature at which the effective potential
exhibits two degenerate minima, one at zero and the other at $\upsilon(T_{c})
$. Both $T_{c}$ and $\upsilon(T_{c})$ are determined using the full effective
potential at finite temperature (\ref{Veff}).

In the SM, the ratio $\upsilon(T_{c})/T_{c}$ is approximately $\left(
2m_{W}^{3}+m_{Z}^{3}\right)  /\left(  \pi\upsilon m_{h}^{2}\right)  $, and
therefore the criterion for a strongly first phase transition is not fulfilled
for $m_{h}>42~\mathrm{GeV}$. However, if the one-loop corrections in
(\ref{mh}) are sizeable, then this bound could be relaxed in such a way that
the Higgs mass is consistent with the measured value at the LHC. This might be
possible since the extra charged scalars affect the dynamics of the SM scalar
field vev around the critical temperature \cite{hna}.

In this model, the one loop correction to the Higgs mass due to the charged
singlets, when neglecting the Higgs and gauge bosons contributions, is%
\begin{equation}
m_{h}^{2}\simeq2\lambda\upsilon^{2}+{\sum\limits_{i}}\frac{\lambda_{i}%
^{2}\upsilon^{2}}{16\pi^{2}}\ln\frac{m_{S_{i}}^{2}}{m_{t}^{2}},\label{mh}%
\end{equation}
where the first term on the right hand side of the equation is the Higgs mass
at the tree level. If one takes $m_{S_{1}}=m_{S_{2}}=2m_{t}$ and $\lambda
_{1}=\lambda_{2}$, then the Higgs mass is exactly 125$\;\mathrm{GeV} $ for
$\lambda=10^{-1}$, $10^{-2}$, $10^{-3}$ if $\lambda_{1}=1.82$, $3.68$, $3.82$,
respectively. Note that these values are still within the perturbative regime.
On the other hand, these extra corrections could be negative and may relax the
large tree-level mass value of the Higgs to its experimental value for
$\lambda$ large. Therefore, it is expected that these extra charged scalars
will help the EWPT to be strongly first order by enhancing the value of the
effective potential at the wrong vacuum at the critical temperature without
suppressing the ratio $\upsilon(T_{c})/T_{c}$, and therefore avoiding the
severe bound on the mass of the SM Higgs. However, as it has been shown in the
previous section, the relic density requires large values for $m_{S_{1}}$ and
so the Higgs mass in Eq. (\ref{mh}) can be easily set to its experimental
value ($125~GeV$),while keeping $S_{2}$ light, for small doublet quartic
coupling (which gives a strong EWPT). Thus, both the measured values of the
Higgs mass and the requirement for the EWPT to be strongly first order are not
in conflict with values of $m_{2}$ smaller than $245\;\mathrm{GeV}$ (as
required from the observed relic density).

During the Universe cooling, the Higgs vev decreases slower than the SM case,
where it decays quickly to zero just around $T\sim100\;\mathrm{GeV}$. Here it
is delayed up to $\mathrm{TeV}$ due to the existence of the extra charged
scalars; and the EWPT occurs around $T\sim100\;\mathrm{GeV}$ due to the fact
that the effective potential at the wrong vacuum ( $<h>=0$) is
temperature-dependant through the charged scalars thermal masses in the
symmetric phase \cite{kntJCAP}.

In July 2012, ATLAS \cite{ATLAS2} and CMS \cite{CMS2} collaborations have
announced the observation of a scalar particle with mass $\simeq125$
\textrm{GeV} at about $5~\sigma$ confidence level. The question is whether or
not this is really the SM Higgs or some Higgs-like state with different
properties. Indeed, the fit of the data by the ATLAS collaboration seems to
show an excess in $h\rightarrow\gamma\gamma$ events by more than $50\%$ with
respect to the SM, while the updated CMS analysis is consistent with the SM.
Defining $R_{\gamma\gamma}$ and $R_{\gamma Z}$\ to be the decay width of
$h\rightarrow\gamma\gamma$ \ and $h\rightarrow\gamma Z$ respectively, scaled
by their expected SM value, we find that
\begin{align}
R_{\gamma\gamma}  &  =\left\vert 1+\frac{\upsilon^{2}}{2}\frac{\frac
{\lambda_{1}}{m_{S_{1}}^{2}}A_{0}\left(  \tau_{S_{1}}\right)  +\frac
{\lambda_{2}}{m_{S_{2}}^{2}}A_{0}\left(  \tau_{S_{2}}\right)  }{A_{1}\left(
\tau_{W}\right)  +N_{c}Q_{t}^{2}A_{1/2}\left(  \tau_{t}\right)  }\right\vert
^{2},\label{Ryy}\\
R_{\gamma Z}  &  =\left\vert 1+s_{w}^{2}\frac{\upsilon^{2}}{2}\frac
{\frac{\lambda_{1}}{m_{S_{1}}^{2}}A_{0}\left(  \tau_{S_{1}},\zeta_{S_{1}%
}\right)  +\frac{\lambda_{2}}{m_{S_{2}}^{2}}A_{0}\left(  \tau_{S_{2}}%
,\zeta_{S_{2}}\right)  }{c_{w}A_{1}\left(  \tau_{W},\zeta_{W}\right)
+\frac{2\left(  1-8s_{w}^{2}/3\right)  }{c_{w}}A_{1/2}\left(  \tau_{t}%
,\zeta_{t}\right)  }\right\vert \label{RyzZ}%
\end{align}
where $\tau_{X}=m_{h}^{2}/4m_{X}^{2}$, $\zeta_{X}=m_{Z}^{2}/4m_{X}^{2}$, with
$m_{X}$ is the mass of the charged particle X running in the loop, $N_{c}=3$
is the color number, $Q_{t}$ is the electric charge of the top quark in unit
of $\left\vert e\right\vert $, and the loop functions $A_{i}$ are defined in
\cite{djouadi}. It is clear that the effect on $B(h\rightarrow\gamma\gamma)$
the charged scalar singlets will depend on how light are $S_{1,2}^{\pm}$, the
sign and the strength of their couplings to the SM Higgs doublet. For
instance, an enhancement can be achieved by taking $\lambda_{1}$ and/or
$\lambda_{2}$ to be negative.

\begin{figure}[t]
\includegraphics[width=7cm,height=5cm]{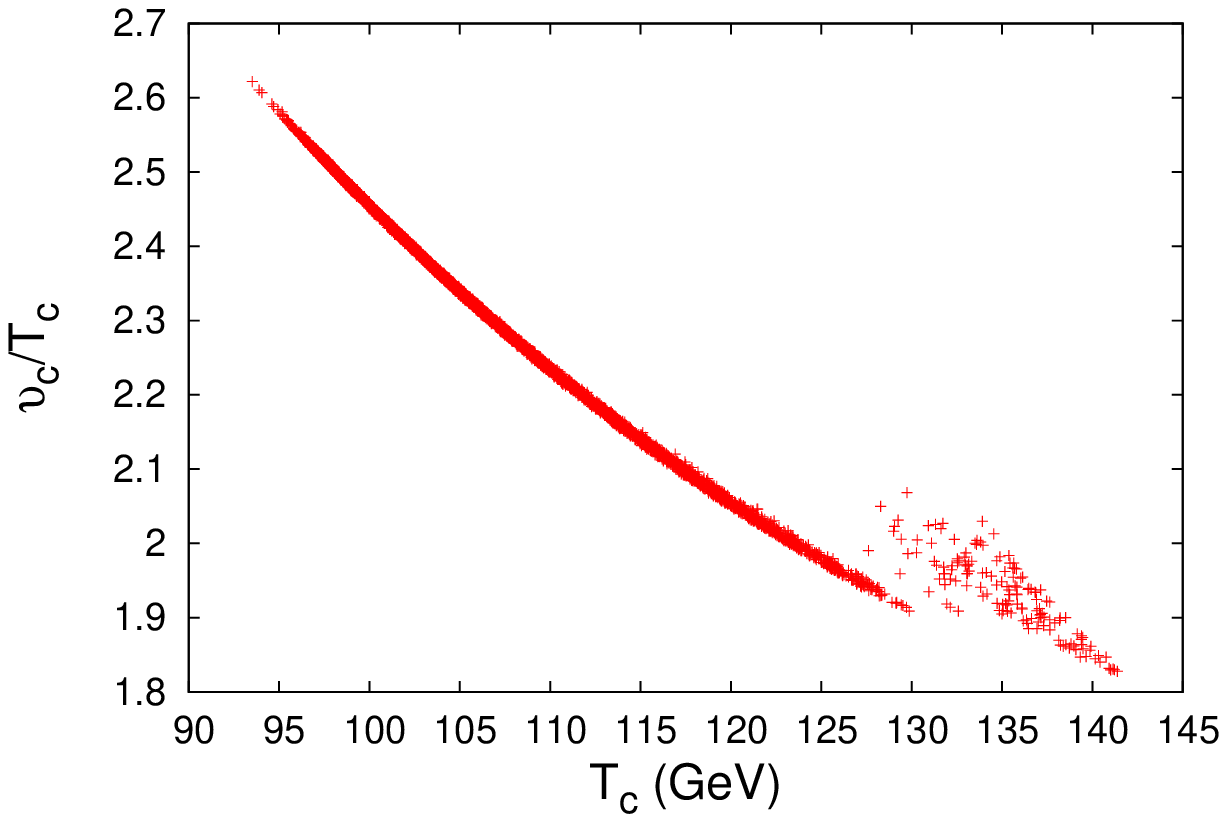}
\includegraphics[width=7cm,height=5cm]{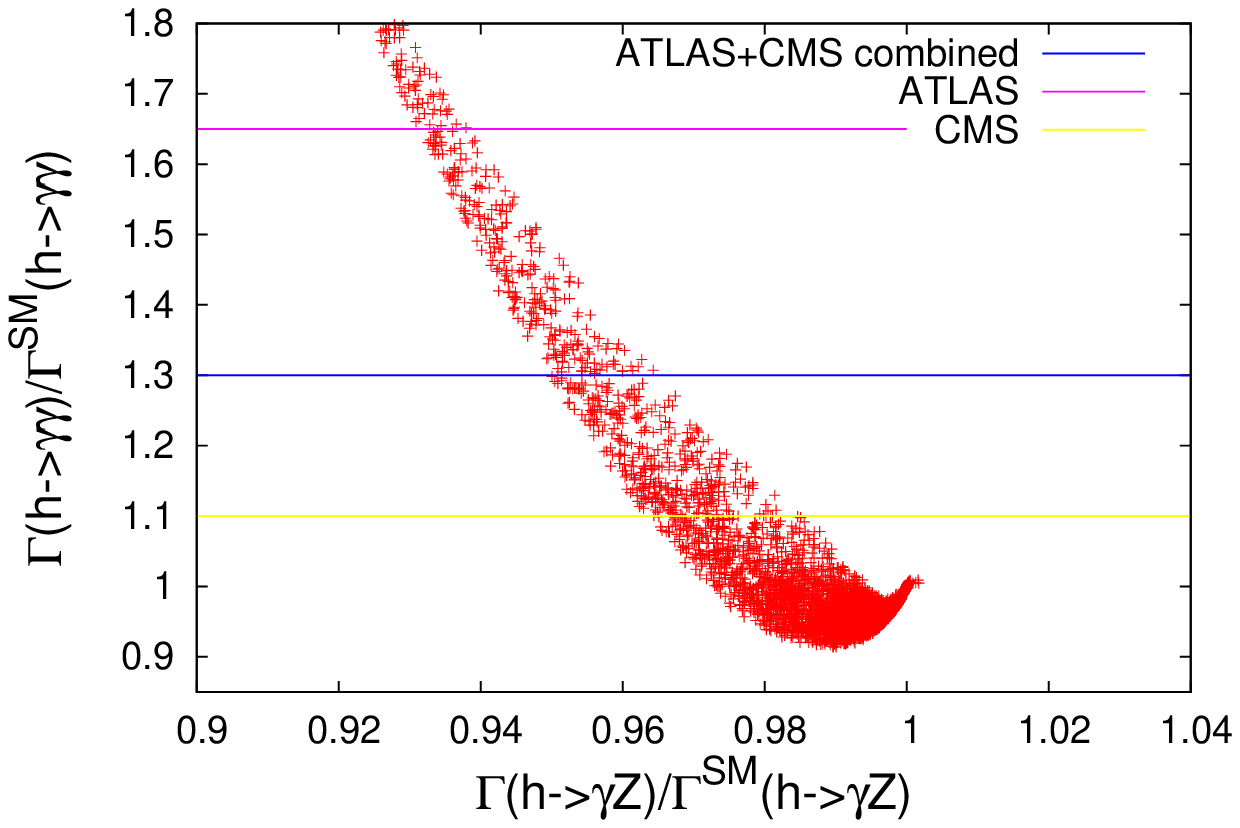}\caption{\textit{Left: In
the left figure, the critical temperature is presented versus the quantity
$\upsilon_{c}/T_{c}$ in (\ref{v/t}). In the right one, the relative
contribution of the one-loop corrections (including the counter-terms) to the
Higgs mass versus the parameter $\lambda$. Right: The modified Higgs decay
rates $B(h\rightarrow\gamma\gamma)$ vs $B(h\rightarrow\gamma Z)$, scaled by
their SM values, due to the extra charged scalars, for randomly chosen sets of
parameters. The magenta (yellow) line represents the ATLAS (CMS) recent
measurements on the $h\rightarrow\gamma\gamma$ channel, while the blue one is
their combined result.}}%
\label{ct}%
\end{figure}

In Fig. \ref{ct}, we present the ratio $\upsilon(T_{c})/T_{c}$\ versus the
critical temperature $T_{c}$ (left); and $R_{\gamma\gamma}$ versus $R_{\gamma
Z}$ (right) for randomly chosen sets of parameters where the charged scalars
are taken to be heavier than 100 $\mathrm{GeV}$, the Higgs mass within the
range $124<m_{h}<126~\mathrm{GeV}$. In our numerical scan, we take the model
parameters relevant for the Higgs decay to be in the range%
\begin{equation}
\lambda<2,~\left\vert \lambda_{1,2}\right\vert <3,~m_{1,2}^{2}<2~\mathrm{TeV}%
^{2},\label{par}%
\end{equation}
where the Higgs mass is calculated at one-loop level. An enhancement of
$B(h\rightarrow\gamma\gamma)$ can be obtained for a large range of parameter
space, whereas $B(h\rightarrow\gamma Z)$ is slightly reduced with respect to
the SM. It is interesting to note that if one consider the combined ATLAS and
CMS di-photon excess, then $R_{\gamma Z}$ is predicted to be smaller than the
expected SM value by approximately $5\%$.

From Fig. \ref{ct}, we can see that one can have a strongly first order EWPT
while the critical temperature lies around 100 $\mathrm{GeV}$ without being in
conflict with the measured value of the Higgs mass; while $R_{\gamma\gamma}$
can have an enhancement where $R_{\gamma Z}$ remains almost constant.

\section{Phenomenology at the ILC}

The RH neutrinos do couple to the charged leptons only, then one excepts them
to be produced at $e^{-}e^{+}$ colliders, such as the ILC and CLIC with a
collision energy $\sqrt{s}$ of few hundreds \textrm{GeV} up to TeV. If a
$N_{i}N_{k}$ pair is produced inside the collider, they will decay to pairs of
charged leptons and $N_{1}N_{1}$, where the charged leptons have not necessary
the same flavor. If such decays occur inside the detector, then the signal
will be
\[
\left\{
\begin{array}
[c]{ccc}%
\not E  & \text{for} & e^{+}e^{-}\rightarrow N_{1}N_{1}\\
\not E  +2\ell_{R}, & \text{for} & e^{+}e^{-}\rightarrow N_{1}N_{2,3}\\
\not E  +4\ell_{R}, & \text{for} & e^{+}e^{-}\rightarrow N_{2,3}N_{2,3},
\end{array}
\right.
\]
and since $m_{N_{i}}\geq100~$\textrm{GeV}, it is very possible that the decay
$N_{2,3}\rightarrow N_{1}+2\ell_{R}$ occurs outside the detector, and thus
escapes the detection, and therefore considered to be missing energy. The
process $e^{+}e^{-}\rightarrow e^{-}\mu^{+}+\not E  $\ can be a good probe for
this model, however there are here so many contributions to the missing energy
whose sources both the f's and g's vertices in (\ref{L}) \cite{ANS}. Here, we
analyze the production of all possible pairs of RH neutrinos, tagged with a
photon from an initial state radiation, that is $e^{-}e^{+}\rightarrow
N_{i}N_{k}\gamma$ (with $i,k=1,2,3$), where one searches for a high $p_{T}$
gamma balancing the invisible RH neutrinos.

If the emitted photon is soft or collinear, then one can use the
soft/collinear factorization form \cite{Birkedal}
\begin{align}
\frac{d\sigma\left(  e^{+}e^{-}\rightarrow N_{i}N_{k}\gamma\right)  }%
{dxd\cos\theta}  &  \simeq\mathcal{F}(x,\cos\theta)\hat{\sigma}\left(
e^{+}e^{-}\rightarrow N_{i}N_{k}\right)  ,\label{fsr}\\
\mathcal{F}(x,\cos\theta)  &  =\frac{\alpha_{em}}{\pi}\frac{1+(1-x)^{2}}%
{x}\frac{1}{\sin^{2}\theta}.
\end{align}
with $x=2E_{\gamma}/\sqrt{s}$, here $\theta$\ is the angle between
the photon and electron and $\hat{\sigma}$ is the cross section
evaluated at the reduced center of mass energy $\hat{s}=(1-x)s$.
Collinear photon with the incident electron or positron could be a
good positive signal, especially if the enhancement in (\ref{fsr})
is more significant than the SM background.

There are two leading SM background processes: a) the neutrino counting
process $e^{-}e^{+}\rightarrow\nu_{i}{\bar{\nu}}_{i}\gamma$ from the t-channel
$W$ exchange and the s-channel $Z$ exchange, and b) the Bhabha scattering with
an extra photon $e^{-}e^{+}\rightarrow e^{-}e^{+}\gamma$, which can mimic the
$N_{i}N_{i}$ signature when the accompanying electrons or photons leave the
detector through the beam pipe \cite{BOUJ}. In addition to putting the cut on
the energy of the emitted photon, one can reduce further the mono-photon
neutrino background, by polarizing the incident electron and positron beams
such that
\begin{equation}
\frac{N_{e_{R}^{-}}-N_{e_{L}^{-}}}{N_{e_{R}^{-}}+N_{e_{L}^{-}}}%
>>50\%;~~~~~~~\frac{N_{e_{R}^{+}}-N_{e_{L}^{+}}}{N_{e_{R}^{+}}+N_{e_{L}^{+}}%
}<<50\%,
\end{equation}
where $N_{e_{L,R}^{-}}$ and $N_{e_{L,R}^{+}}$ are the number densities of the
left (right)-handed electrons and positrons per unit time in the beam. At
$\sqrt{s}>>100$ \textrm{GeV} the process $e^{-}e^{+}\rightarrow\nu_{i}%
{\bar{\nu}}_{i}\gamma$ is dominated by the $W$-exchange, and hence one expect
that having the electron (positron) beam composed mostly of polarized right
handed (left handed) electrons (positron) reduces this background
substantially, whereas the signal increases since $N_{i}$ couples to the right
handed electrons.

When estimating the total cross section $\sigma\left(  e^{+}e^{-}\rightarrow
N_{i}N_{k}\right)  $ \cite{kntJCAP}, we can present the differential cross
section versus the photon energy for two values of collision energies
$\sqrt{s}=500$\ \textrm{GeV}\ and 1 \textrm{TeV}; using the approximation
(\ref{fsr}), by integrating over the angle $\theta$ taking into account the
minimum value of electromagnetic calorimeter acceptance in the ILC to be
$\sin\theta>0.1$\ \cite{ILC}.

\begin{figure}[t]
\begin{centering}
\includegraphics[width=7cm,height=5cm]{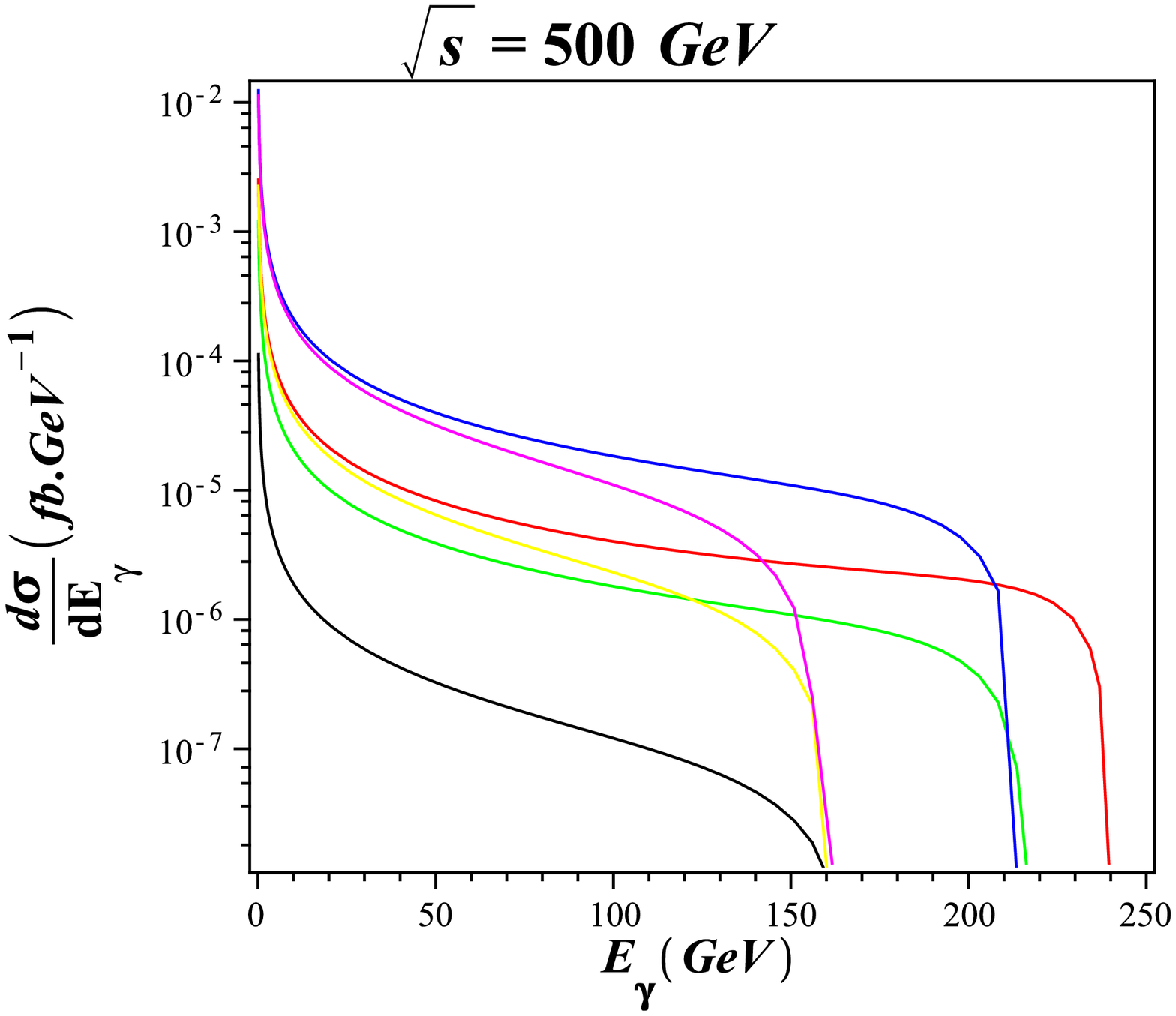}\includegraphics[width=7cm,height=5cm]{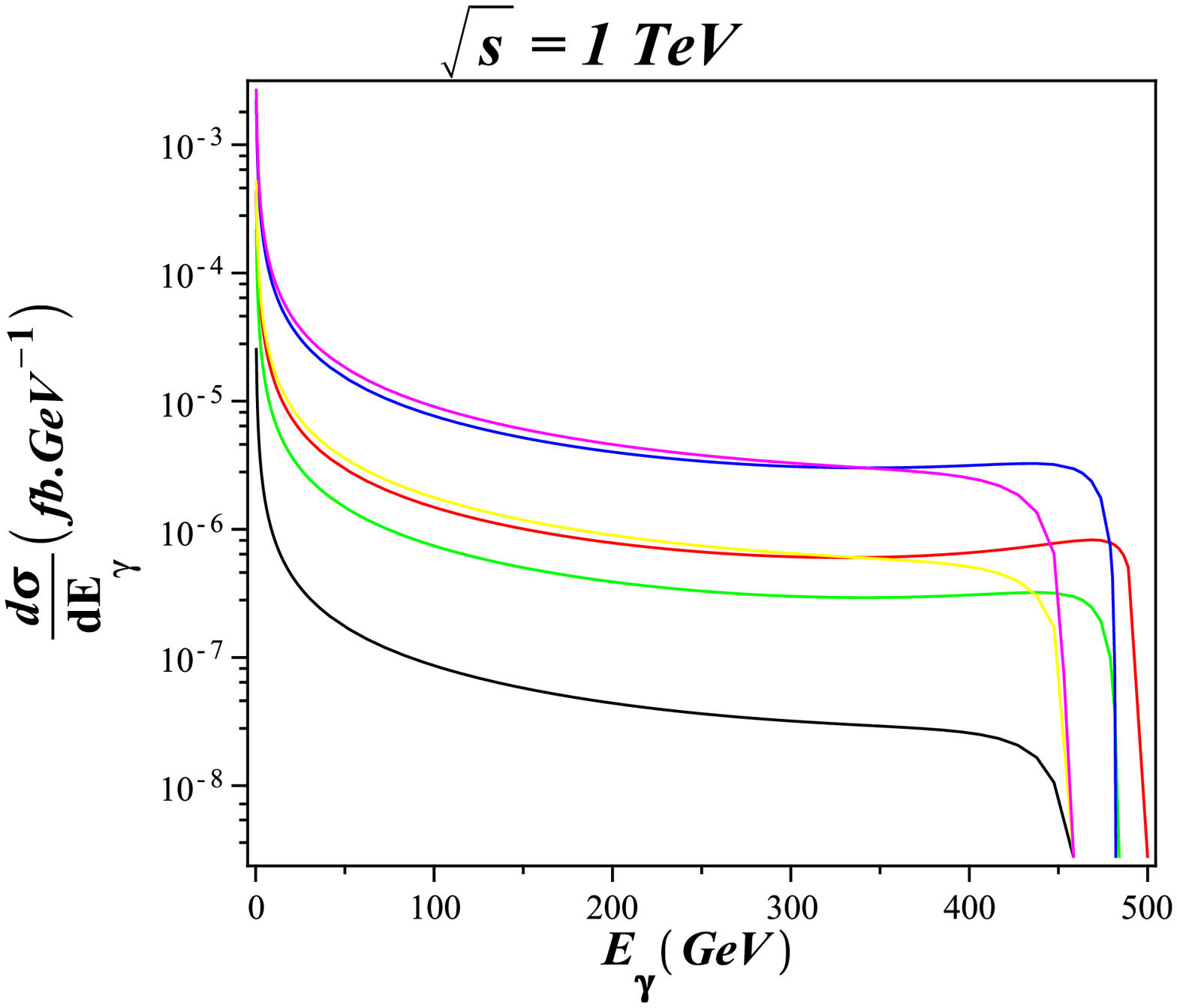}
\par\end{centering}
\caption{\textit{The photon spectra from the processes $e^{+}e^{-}\rightarrow
N_{i}N_{k}\gamma$ where the curves: {red, green, black, blue, yellow, magenta}
correspond to (i,k)={(1,1), (1,2), (2,2), (1,3), (2,3), (3,3)} respectively.
Here, we considered the following favored mass values: $m_{N_{1}}=52.53$ GeV,
$m_{N_{2}}=121.80$ GeV, $m_{N_{3}}=126.19$ GeV, $m_{S_{2}}=144.28$ GeV, and
the coupling values: $g_{1e}=-4.19\times10^{-2}$, $g_{2e}=2.10\times10^{-2}$
and $g_{3e}=-6.75\times10^{-2}$.}}%
\label{Ga}%
\end{figure}We see that for the benchmark shown in Fig. \ref{Ga}, the heaviest
RH neutrino is largely produced due to its large couplings to the
electron/positron. Thus, for this particular benchmark the missing energy is
dominated not by the DM, but rather by the other RH neutrinos. Another
interesting process that might be possible to search for at both lepton and
hadron colliders is the production of $S_{1,2}^{\pm}$. At the LHC they can be
pair produced in an equal number via the Drell-Yan process, with a production
rate of $S_{1,2}^{\pm}$ that is suppressed at very high energies, and so we
expect that most of the produced $S_{1,2}^{\pm}$ will have energies not too
far from their masses. Then, each pair of charged scalars decays into charged
leptons and missing energy, such as $e^{+}e^{-}$, $\mu^{+}\mu^{-}$, $\mu
^{+}e^{-}$. The observation of an electron (positron) and anti-muon (muon),
will be a strong signal for the production of the charged scalars of this
model. The energy carried out by the charged leptons, $\ell_{\alpha}^{+}%
\ell_{\beta}^{-}$, produced in the decay of $S_{1,2}^{\pm}$ will be limited by
the phase space available to $N_{1}$ and $\ell_{\alpha,\beta}$ since
$m_{S_{2}}-m_{N_{1}}<<m_{S_{2}}$. On the other hand, the leptons originating
from the decay of $S_{1}^{\pm}$ will be produced in association with a SM
neutrino, and hence can have energy as large as $\sim m_{S_{1}}$. Thus, by
putting the appropriate energy cuts on the energy of final states $e^{\pm}%
\mu^{\mp}$ and discriminating the SM background (from the decay of
$pp\rightarrow W^{+}W^{-}+X\rightarrow\ell_{\alpha}\ell_{\beta}+\nu^{\prime}%
s$), one can, in principle, identify the signal for the charged scalars. This
requires a detailed a analysis which we plan to carry out in a future
publication \cite{ANS}.

\section{Conclusion}

In this paper, we analyzed a radiative model for neutrino masses, generated at
three loop level. Beside it can accommodate the neutrino oscillation data and
be consistent with the LFV processes, it provides a DM candidate with a mass
lying between few \textrm{GeV} up to 225 \textrm{GeV}; and a relatively light
charged scalar, $S_{2}^{\pm}$, with a mass below 245 \textrm{GeV}.
Furthermore, we showed that the charged scalar singlets can give an
enhancement for $B\left(  h\rightarrow\gamma\gamma\right)  $, whereas the
decay $B\left(  h\rightarrow\gamma Z\right)  $ get a small suppression,
compared to the SM. We also found that charged scalars with masses close the
electroweak scale make the electroweak phase transition strongly first order.
Since $N_{1}$ couples only to leptons, it can not be observed in experiments
for direct dark matter searches. However it might be possible to search for
such particle in indirect detection experiments, such as Fermi-LAT, and at
future linear colliders, such as the international linear collider (ILC).
Thus, for this particular benchmark the missing energy is dominated not by the
DM, but rather by the other RH neutrinos.

\bigskip

\textbf{Acknowledgments}: This work is supported by the Algerian Ministry of
Higher Education and Scientific Research under the PNR '\textit{Particle
Physics/Cosmology: the interface}', and the CNEPRU Project No.
\textit{D01720130042}.


\begin{thebibliography}{99}                                                                                               %
\bibitem {seesaw}P. Minkowski, Phys. Lett. B 67 (1977) 421; M. Gell-Mann, P.
Ramond and R. Slansky, Proceedings of the Supergravity Stony Brook
Workshop, New York 1979, eds. P. Van Nieuwenhuizen and D. Freedman;
T. Yanagida, Proceedings of the Workshop on Unified Theories and
Baryon Number in the Universe, Tsukuba, Japan 1979, eds. A. Sawada
and A. Sugamoto; R. N. Mohapatra and G. Senjanovic, Phys. Rev. Lett.
44 (1980) 912; J. Schechter and J.W.F. Valle, Phys. Rev. D 22 (1980)
2227.

\bibitem {FY}M. Fukugita and T. Yanagida, Phys. Lett. B 174 (1986) 45.

\bibitem {Zee}A. Zee, Phys. Lett. B 161 (1985) 141.

\bibitem {solnu}X.-G. He, Eur. Phys. J. C 34 (2004) 371; P.H. Frampton,
M.C. Oh and T. Yoshikawa, Phys. Rev. D 65 (2002) 073014.

\bibitem {Babu}K.S. Babu, Phys. Lett. B 203 (1988) 132.

\bibitem {KNT}L.M. Krauss, S. Nasri and M. Trodden, Phys. Rev. D 67 (2003) 085002.

\bibitem {AKS}M. Aoki, S. Kanemura and O. Seto, Phys. Rev. Lett. 102
(2009) 051805; Phys. Rev. D 80 (2009) 033007; M. Aoki, S. Kanemura,
K. Yagyu, Phys. Rev. D 83 (2011) 075016; M. Gustafsson, J.M. No and
M.A. Rivera,     Phys. Rev. Lett. 110 (2013) 211802.

\bibitem {Keung}K. Cheung and O. Seto, Phys. Rev. D 69 (2004) 113009.

\bibitem {kntJCAP}A. Ahriche and S. Nasri, JCAP 07 (2013) 035.

\bibitem {LFV}J. Adam et al. [MEG Collaboration], arXiv:1303.0754 [hep-ex].

\bibitem {pdg}J. Beringer et al. (Particle Data Group), Phys. Rev. D 86 (2012)
010001.

\bibitem {ATLAS}G. Aad et al. (ATLAS Collaboration), Phys. Lett. B 716 (2012) 1.

\bibitem {CMS}S. Chatrchyan et al. (CMS Collaboration), Phys. Lett. B 716 (2012) 30.

\bibitem {PMNS}B. Pontecorvo, Sov. Phys. JETP 26 (1968) 984; Z. Maki, M.
Nakagawa and S. Sakata, Prog. Theor. Phys. 28 (1962) 870.

\bibitem {GF}D.V. Forero, M. Tortola and J.W.F. Valle, Phys. Rev. D 86 (2012) 073012.

\bibitem {bbo}F. Simkovic, A. Faessler, H. Muther, V. Rodin and M. Stauf,
Phys. Rev. C 79 (2009) 055501.

\bibitem {Planck}P.A.R. Ade et al. [Planck Collaboration], arXiv:1303.5062 [astro-ph.CO].

\bibitem {Th}L. Dolan and R. Jackiw, Phys. Rev. D 9 (1974) 3320; S.
Weinberg, Phys. Rev. D 9 (1974) 3357.

\bibitem {ring}M.E. Carrington, Phys. Rev. D 45 (1992) 2933.

\bibitem {EWB}V. Kuzmin, V. Rubakov and M.E. Shaposhnikov, Phys. Lett. B 155 (1985)
36.

\bibitem {SFOPT}M.E. Shaposhnikov, Nucl. Phys. B 287 (1987) 757; ibid
299 (1988) 797.

\bibitem {hna}A. Ahriche, Phys. Rev. D 75 (2007) 083522; A. Ahriche and S.
Nasri, Phys. Rev. D 83 (2011) 045032; Phys. Rev. D 85 (2012) 093007.

\bibitem {ATLAS2}ATLAS Collaboration, Report No. ATLAS-CONF-2013-014, March 2013.

\bibitem {CMS2}C. Ochando [CMS Collaboration], Talk presented at Rencontres de
Moriond, La Thuile, Italy, 9-16, March 2013.

\bibitem {djouadi}A. Djouadi, Phys. Rept. 457 (2008) 1.

\bibitem {ANS}A. Ahriche, S. Nasri and R. Soualah, in preparation.

\bibitem {Birkedal}A. Birkedal, K. Matchev and M. Perelstein, Phys. Rev. D 70 (2004)
077701.

\bibitem {BOUJ}C. Bartels, M. Berggren and J. List, Eur. Phys. J. C 72 (2012) 2213.

\bibitem {ILC}J.A. Aguilar-Saavedra et al. [ECFA/DESY LC Physics Working Group
Collaboration], hep-ph/0106315.
\end{thebibliography}
\end{document}